\documentstyle[prb,aps,twocolumn]{revtex}
\begin{document}
\draft
\title{The re-examined phenomenological phase transitions theory for ferromagnets}
\author{V.A.Skrebnev}
\address{Physics Department, Kazan State University, 420008 Kazan, Russia}
\maketitle
\date{\today }

\begin{abstract}
The existence of the linear on the order parameter term of the thermodynamic
functions expansion near the critical point is justified. The criticism of
the arguments, used for the rejection of the odd-power expansion terms of
the ferromagnets thermodynamic functions is presented. It is shown, that
taking into account the linear term in expansion one achieves the
consentency with experimental data on the magnetization behavior near the
transition temperature in ferromagnets.

PACS nimbers: 75.40.-s, 75.40.Cx
\end{abstract}


\section{Introduction}

Experimental investigations of the critical phenomena show, that the Landau
phase transitions theory does not agree with an experiment. Usually, this
fact is attributed to the large fluctuations near the critical point.
However the Landau conclusions contradict even the experimental data, which
are received as an exact measurements result (fluctuations are proved
negligible). The experiments of Heller and Benedek \cite{Heller} may be the
example. In these experiments the temperature dependence of the
magnetization near the critical point has been studied. The measurements
were carried out by the nuclear magnetic resonance method. In the case of
large fluctuations the measurements of the magnetization would be
impossible. Therefore, the explanation of discrepancy of the theory and the
experiment \cite{Heller} by very large fluctuations seems to be not
convincing enough.

The assumptions which lie in the basis of the Landau theory look natural and
simple. That is why we decide to re-analyze carefully this theory. In the
process of our analysis we have found one essential circumstance, with which
it is necessary to acquaint the reader.

\section{The modification of the Landau theory}

By examining the symmetry change at the second type phase transitions,
Landau presented the crystal density function in the form: 
\begin{equation}
\rho =\sum_{i,n}\eta _i^{(n)}\psi _i^{(n)}.  \label{eq1}
\end{equation}
Here $n$ is the index of the irreducible representation of the crystal
symmetry group $G$ in the high symmetry phase and $\psi _i^{(n)}$ are the
basic functions of these irreducible representations.

Denoting by $\rho _0$ the invariant at all transformations of the $G$ group
function (this function realizes the unit representation of the $G$ group),
Landau wrote: 
\begin{equation}
\rho =\rho _0+\delta \rho ^{\prime },  \label{eq2}
\end{equation}
where 
\begin{equation}
\delta \rho ^{\prime }=\sum_{i,n}^{\prime }\eta _i^{(n)}\psi _i^{(n)},
\label{eq3}
\end{equation}
and the unit representation ($n=1$) is excluded from the summation.

Near the phase transition temperature (critical point $T_c$) the expansion
of thermodynamic functions in the Landau theory is realized on the powers of
small $\delta \rho ^{\prime }$ with $\rho _0$ kept invariable. We suggest
that the contribution to the crystal density of the function $\rho _0$ can
not remain equal to $\rho _0\left( T_c\right) $ as the temperature changes.
Otherwise, far from the critical point, $\rho _0\left( T\right) $ would be
equal to $\rho _0\left( T_c\right) $, which is obviously not the case.

Instead, it is naturally to present the crystal density $\rho $ as 
\begin{equation}
\rho =\rho _0\left( T_c\right) +\delta \rho ,  \label{eq4}
\end{equation}
\begin{equation}
\delta \rho =\delta \rho _0+\delta \rho ^{\prime },  \label{eq5}
\end{equation}
\begin{equation}
\delta \rho _0=\eta ^{(1)}\psi ^{(1)}.  \label{eq6}
\end{equation}
We believe that the expansion of thermodynamic functions on powers of $%
\delta \rho $ is mathematically and physically more correct, than the
expansion on powers of $\delta \rho ^{\prime }$ (\ref{eq3}). We verify this
conjecture in the next Section by calculting the critical index for
magnetization and general thermodynamic relations for ferromagnets.

The invariants of the second and higher orders in the expansion of the
thermodynamic functions near the critical point correspond to the density
change $\delta \rho ^{\prime }$, which does not consist the unit
representation. In particular, the second order invariant has the form: 
\begin{equation}
\eta ^2=\sum_i\eta _i^2,  \label{eq7}
\end{equation}
with $\eta $ being the quantitative measure of the deviation from the
critical point.

The linear invariant $\eta ^{(1)}$ corresponds to the density change $\delta
\rho _0$, which transforms according to the unit representation. This
invariant does not determine the symmetry change and does not independent.
The magnitudes $\delta \rho _0$ and $\delta \rho ^{\prime }$ are of the same
order. Hence, $\eta ^{(1)}$ is proportional to $\eta $. This means, that in
the expansion of the thermodynamic functions the linear on $\eta $ term
presents. Below we will show for ferromagnets, that keeping the linear term
provides the consistency of experiment and theory.

\section{The linear term and the critical phenomena in ferromagnets}

Following Landau the linear terms of the thermodynamic potentials expansion
are rejected in description of the critical phenomena. One usually uses some
additional arguments to exclude the odd terms in the expansion of the
ferromagnet's thermodynamic functions. In the book \cite{Belov} it is
claimed, that the scalar function expansion on the vector quantity may only
contain the even power of this quantity. However it is not difficult to
show, that this statement is incorrect. Indeed, the first law of
thermodynamics for the magnetic systems can be written as 
\begin{equation}
dU=TdS+{\bf H}d{\bf M.}  \label{eq8_1}
\end{equation}
For the Helmholtz potential $A(T,M)$ we have: 
\begin{equation}
dA=-SdT+{\bf H}d{\bf M.}  \label{eq8}
\end{equation}
From the expression (\ref{eq8})\ it follows that 
\begin{equation}
{\bf H}=\frac{\partial A}{\partial {\bf M}}={\bf n}\frac{\partial A}{%
\partial M},  \label{eq9}
\end{equation}
where ${\bf n}$ is the unit vector along the ${\bf M}$ direction. Thus, we
can rewrite Eq. (\ref{eq8}) in the following way: 
\begin{equation}
dA=-SdT+HdM.  \label{eq10}
\end{equation}
The example illustrates the general situation that only numerical
characteristics of the vectors do appear (via the scalar products) in the
expression for the thermodynamic functions. Therefore, it is instructive to
expand the ferromagnet's thermodynamic functions on powers of magnetic
moment magnitude. Thus, it is not possible to reject the terms of the
expansion with odd powers of the magnetic moment magnitude declaring that
the magnetic moment is vector.

In the book \cite{Stanley} the absence of the $M$ odd powers in the
ferromagnet's thermodynamic functions expansion is justified by the
statement, that these functions are even regarding $M$. However, the change
in the $M$ sign in a ferromagnet is confined to the change of the magnetic
field sign (see Eq. (\ref{eq9})). At the simultaneous change of $M$ and $H$
signs the thermodynamic functions values do not change. If we expand the
thermodynamic function on the magnetic moment magnitude, when $M$ changes
sign, the non-zero coefficients at odd $M$ powers also change the sign, and
the independence of the thermodynamic function on the $M$ sign will be
ensured.

Let us expand the potential $A(T,M)$ up to fourth power on $M$ near the
critical point: 
\begin{equation}
A(T,M)=\sum_{n=0}^4L_n(T)M^n.  \label{eq11}
\end{equation}
For the ferromagnetic phase the equilibrium value of $M$ is determined from
the expression: 
\begin{eqnarray}
H &=&\left( \frac{\partial A}{\partial M}\right)  \nonumber  \label{eq15} \\
\ &=&L_1(T)+2L_2(T)M+3L_3(T)M^2+4L_4(T)M^3.  \label{eq15}
\end{eqnarray}
Then consider separately the term $L_1(T)$ of equation (\ref{eq15}). The
expansion of $L_1(T)$ on $t=T-T_c$ powers up to the first power has the
form: 
\begin{equation}
L_1(T)=L_1(T_c)+t\left( \frac{\partial L_1}{\partial T}\right) _{T_c}.
\label{eq12}
\end{equation}
In the ferromagnetic phase $L_1(T)=0$, since at $T=T_c$ the equilibrium
value $M=0$ in the case $H=0$. Hence, the coefficient $L_1(t)$ is given by 
\begin{equation}
L_1(T)=at,  \label{eq13}
\end{equation}
where 
\begin{equation}
a=\left( \frac{\partial L_1}{\partial T}\right) _{T_c}=\left( \frac{\partial
^2A}{\partial T\partial M}\right) _{T_c}=\left( \frac{\partial H}{\partial T}%
\right) _{T_c}.  \label{eq14}
\end{equation}
Therefore, the coefficient $L_1$ changes the sign at the $M$ sign change,
that is confined with the $H$ sign change. Thus, the rejection of the linear
term of the expansion (\ref{eq11}) has no serious theoretical reasons.

Expanding the coefficients $L$ on powers of $t$ , we rewrite Eq. (\ref{eq15}%
) in the form: 
\begin{equation}
H=\sum_{m,n}^{m+n\leq 3}a_{mn}t^mM^n.  \label{eq16}
\end{equation}
For the magnetic systems at the phase transition point we have the following
relations (in accordance with the general theory of the second-type phase
transitions): 
\begin{eqnarray}
\left( \frac{\partial H}{\partial M}\right) _{T_c} &=&\left( \frac{\partial
^2A}{\partial M^2}\right) _{T_c}=0,  \nonumber  \label{eq17} \\
\left( \frac{\partial ^2H}{\partial M^2}\right) _{T_c} &=&\left( \frac{%
\partial ^3A}{\partial M^3}\right) _{T_c}=0,  \label{eq17} \\
\left( \frac{\partial ^3H}{\partial M^3}\right) _{T_c} &=&\left( \frac{%
\partial ^4A}{\partial M^4}\right) _{T_c}>0.  \nonumber
\end{eqnarray}
Hence, the terms with $M$ and $M^2$ must be absent in Eq. (\ref{eq16}). The
terms proportional to $tM$, $t^2$, $t^2M$, $tM^2$ and $t^3$ are smaller than
the term $at$, and we may neglect these terms. At the same time we must keep
the terms with $M^3$, since {\it a priory} the relative values of $t$ and $M$
unknown. As a result, in the case $H=0$ the equation (\ref{eq16}) takes the
form: 
\begin{equation}
H=at+cM^3=0,  \label{eq18}
\end{equation}
where 
\begin{equation}
c=\frac 16\left( \frac{\partial ^4A}{\partial M^4}\right) _{T_c}=\frac 16%
\left( \frac{\partial ^3H}{\partial M^3}\right) _{T_c}.  \label{eq19}
\end{equation}
From Eq.(\ref{eq18}) we easily find: 
\begin{equation}
M=\left( -\frac{at}c\right) ^\beta ,\,\,\,\,\,\beta =\frac 13.  \label{eq20}
\end{equation}

In the experiments of Heller and Benedek\cite{Heller}
the dependence of $M^3$ on $t$ in MnF$_2$ in zero external
field was studied. This dependence occures to be linear. Thus, taking into
account the linear term in the expansion (12) one achieves the good agreement
with the experiments in zero field.

If the odd terms in the expansion (\ref{eq11}) are rejected, we return to
the Landau-type theory, and get instead of Eq. (\ref{eq18}) 
\begin{equation}
H=btM+cM^3=0,  \label{eq21}
\end{equation}
where 
\begin{equation}
b=\left( \frac{\partial ^3A}{\partial M^2\partial T}\right) _{T_c}=\left( 
\frac{\partial ^2H}{\partial M\partial T}\right) _{T_c}.  \label{eq22}
\end{equation}
From Eq. (\ref{eq21}) it follows: 
\begin{equation}
M=\left( -\frac{bt}c\right) ^\beta ,\,\,\,\,\,\beta =\frac 12,  \label{eq23}
\end{equation}
which contradicts with experiments. Moreover, in the Landau-type theory one
of the equilibrium values of $M$ is zero. At the discussion of this fact it
is affirmed, that zero solution corresponds to the temperature, which is
higher than the Curie point. This statement seems internally inconsistent
with physical meaning of the equation (\ref{eq21}), for which both zero and
non-zero solutions correspond to the same temperature. Alternatively, in our
theory the spurious, non-physical solution, $M=0$, does not appear.

It is known, that for magnetic systems the correlation must be fulfilled 
\cite{Stanley}: 
\begin{equation}
-\left( \frac{\partial M}{\partial T}\right) _H\cdot \left( \frac{\partial H%
}{\partial M}\right) _T=\left( \frac{\partial H}{\partial T}\right) _M.
\label{eq24}
\end{equation}
Using Eq. (\ref{eq18}) and Eq. (\ref{eq20}), we find at $T$ = $T_c$ in
correspondence with relation (\ref{eq24}): 
\begin{equation}
-\left( \frac{\partial M}{\partial T}\right) _H\cdot \left( \frac{\partial H%
}{\partial M}\right) _T=a=\left( \frac{\partial H}{\partial T}\right) _M.
\label{eq25}
\end{equation}
If linear term in the expansion is rejected, from Eq.(\ref{eq21}) and Eq. (%
\ref{eq23}) at $T=T_c$ we find: 
\begin{equation}
-\left( \frac{\partial M}{\partial T}\right) _H\left( \frac{\partial H}{%
\partial M}\right) _T=t^{\frac 12}\left( \frac{\partial ^2H}{\partial
M\partial T}\right) ^{\frac 32}\left[ -\frac 6{\left( \frac{\partial ^3H}{%
\partial M^3}\right) }\right] ^{\frac 12}=0.  \label{eq26}
\end{equation}
Hence, the relation (\ref{eq24}) does not hold, that is incompatible with
thermodynamics of magnetic systems.

In the presence of the external field we rewrite Eq.(\ref{eq16}) in the
form: 
\begin{equation}
H=at+btM+cM^3.  \label{eq27}
\end{equation}
We retain in the above equation the term proportional to $tM$, since this
term is essential for the explanation of the critical phenomena in the
strong magnetic fields.

The dependence of the magnetic moment on the external field near the
critical point was studyed in the work.\cite{Belov1} In strong magnetic fields the
dependence of $H/M$ on $M^2$ occured to be linear. This result is in
agreement with Eq. (\ref{eq27}). Indeed, this equation can be converted to
the form: 
\begin{equation}
\frac HM\left( 1-\frac{at}H\right) =bt+cM^2.  \label{eq28}
\end{equation}
In the strong field we can neglect the term $at/H$ in the left-hand part of
Eq. (\ref{eq28}) in comparison with unity. As a result we obtain the linear
dependence of $H/M$ on $M^2$.

In the Landau-type theory the term $at/H$ is absent from the very beginning.
That is why the experiments \cite{Belov1} were considered as the
confirmation of the Landau-type theory.

With the decrease of the field the domain structure is starting to influence
the dependence of $H/M$ on $M^2$. Therefore it is impossible to pick out the
contribution of the $at/H$ term of Eq. (\ref{eq28}) to the above
experimental results.

We may conclude, that the taking into account the linear term in the
expansion of thermodynamic functions is consistent with the experiment both
in the strong and zero magnetic fields.

\section{Conclusion}

We have the serious reasons to consider, that the thermodynamic function
expansion up to the fourth power in order parameter is correct at least for
the three dimension systems. The origin of the Landau-type theory failures
is connected not with the ideological basis of this theory, but with
incorrect disregard of the linear term in thermodynamic functions expansion.
Keeping of the linear term restores consistency of the theory with an
experiment and may promote the better comprehension of phenomena near the
critical point.

\end{document}